\documentclass[10pt,conference]{IEEEtran}
\IEEEoverridecommandlockouts
\usepackage{booktabs} 
\usepackage{color}
\usepackage{multirow}
\usepackage{url}
\usepackage{listings}
\usepackage{scalefnt}
\usepackage{graphicx}
\usepackage{todonotes}

\usepackage[skip=3pt]{caption}

\usepackage[ruled,linesnumbered]{algorithm2e} 

\SetCommentSty{mycommfont}
\SetAlFnt{\small}
\SetAlCapFnt{\small}
\SetAlCapNameFnt{\small}
\SetAlCapHSkip{0pt}
\IncMargin{0.5em}
\SetDataSty{it}

\usepackage{float}
\usepackage{tikz}
\usetikzlibrary{shadows,arrows,positioning,automata}

\usetikzlibrary{shapes}
\usetikzlibrary{patterns}
\usetikzlibrary{fit,calc}

\usepackage{pifont}
\usepackage{epstopdf}
\usepackage{subcaption}
\usepackage{booktabs}
\usepackage[normalem]{ulem}
\usepackage{scalefnt}
\usepackage{courier}

\newtheorem{mydef}{Definition}

\definecolor{amber(sae/ece)}{rgb}{1.0, 0.49, 0.0}
\definecolor{javagreen}{rgb}{0.25,0.5,0.35} %
\definecolor{alizarin}{rgb}{0.82, 0.1, 0.26}
\definecolor{cadmiumgreen}{rgb}{0.0, 0.42, 0.24}
\definecolor{mikadoyellow}{rgb}{1.0, 0.77, 0.05}
\definecolor{azure}{rgb}{0.0, 0.5, 1.0}
\definecolor{cambridgeblue}{rgb}{0.64, 0.76, 0.68}
\definecolor{cadetgrey}{rgb}{0.57, 0.64, 0.69}
\definecolor{coolblack}{rgb}{0.0, 0.18, 0.39}
\definecolor{javared}{rgb}{0.6,0,0} 
\definecolor{javagreen}{rgb}{0.25,0.5,0.35} 
\definecolor{javapurple}{rgb}{0.5,0,0.35} 
\definecolor{javadocblue}{rgb}{0.25,0.35,0.75} 
\definecolor{cinnabar}{rgb}{0.89, 0.26, 0.2}
\definecolor{brightgreen}{rgb}{0.4, 1.0, 0.0}
\definecolor{blue(pigment)}{rgb}{0.2, 0.2, 0.6}
\definecolor{whitesmoke}{rgb}{0.99, 0.99, 0.99}
\definecolor{bisque}{rgb}{1.0, 0.89, 0.77}
\definecolor{blanchedalmond}{rgb}{1.0, 0.92, 0.8}
\definecolor{burlywood}{rgb}{0.87, 0.72, 0.53}
\definecolor{cornellred}{rgb}{0.7, 0.11, 0.11}
\definecolor{britishracinggreen}{rgb}{0.0, 0.26, 0.15}
\lstdefinelanguage{diff}{
	sensitive=true,
	morecomment=[f][\color{gray}][0]{diff},
	morecomment=[f][\color{gray}][0]{index},
	morecomment=[f][\color{blue}][0]{@@},
	morecomment=[f][\color{magenta}][0]{***},
	morecomment=[f][\color{violet}][0]{!},
	morecomment=[f][\color{red!60!black}][0]-,
	morecomment=[f][\color{green!60!black}][0]+,
	morecomment=[f][\color{magenta}][0]{---},
	morecomment=[f][\color{magenta}][0]{+++},
	morecomment=[f][\color{gray}][0]{Binary},
	morecomment=[f][\color{gray}][0]{Only},
	morecomment=[f][\color{gray}][0]{old},
	morecomment=[f][\color{gray}][0]{new},
	morecomment=[f][\color{gray}][0]{rename},
	morecomment=[f][\color{gray}][0]{similarity},
	morecomment=[f][\color{gray}][0]{deleted},
	morecomment=[f][\color{magenta}][0]{***************},
	morecomment=[f][\color{red!60!black}][0]<,
	morecomment=[f][\color{green!60!black}][0]>,
	morecomment=[f][\color{blue}][0]{0},
	morecomment=[f][\color{blue}][0]{1},
	morecomment=[f][\color{blue}][0]{2},
	morecomment=[f][\color{blue}][0]{3},
	morecomment=[f][\color{blue}][0]{4},
	morecomment=[f][\color{blue}][0]{5},
	morecomment=[f][\color{blue}][0]{6},
	morecomment=[f][\color{blue}][0]{7},
	morecomment=[f][\color{blue}][0]{8},
	morecomment=[f][\color{blue}][0]{9},
}[comments]

\newcommand{\inlinecode}[1]{\lstinline[keepspaces=true,breaklines]$#1$}

\newcommand{\etal}{et~al.\@\xspace}

\newcommand{\figref}[1]{Figure~\ref{fig:#1}}

\pgfdeclarelayer{background}
\pgfdeclarelayer{foreground}
\pgfsetlayers{background,main,foreground}

\newlength{\hatchspread}
\newlength{\hatchthickness}
\newlength{\hatchshift}
\newcommand{\hatchcolor}{}
\tikzset{hatchspread/.code={\setlength{\hatchspread}{#1}},
	hatchthickness/.code={\setlength{\hatchthickness}{#1}},
	hatchshift/.code={\setlength{\hatchshift}{#1}},
	hatchcolor/.code={\renewcommand{\hatchcolor}{#1}}}
\tikzset{hatchspread=3pt,
	hatchthickness=0.4pt,
	hatchshift=0pt,
	hatchcolor=black}
\pgfdeclarepatternformonly[\hatchspread,\hatchthickness,\hatchshift,\hatchcolor]
{custom north west lines}
{\pgfqpoint{\dimexpr-2\hatchthickness}{\dimexpr-2\hatchthickness}}
{\pgfqpoint{\dimexpr\hatchspread+2\hatchthickness}{\dimexpr\hatchspread+2\hatchthickness}}
{\pgfqpoint{\dimexpr\hatchspread}{\dimexpr\hatchspread}}
{
	\pgfsetlinewidth{\hatchthickness}
	\pgfpathmoveto{\pgfqpoint{0pt}{\dimexpr\hatchspread+\hatchshift}}
	\pgfpathlineto{\pgfqpoint{\dimexpr\hatchspread+0.15pt+\hatchshift}{-0.15pt}}
	\ifdim \hatchshift > 0pt
	\pgfpathmoveto{\pgfqpoint{0pt}{\hatchshift}}
	\pgfpathlineto{\pgfqpoint{\dimexpr0.15pt+\hatchshift}{-0.15pt}}
	\fi
	\pgfsetstrokecolor{\hatchcolor}
	\pgfusepath{stroke}
}

\tikzstyle{callbacktext}=[draw, fill=blue!20, text width=6.0em, text centered,
minimum height=1.5em,drop shadow]
\tikzstyle{callbackcfgtitle} = [above, text width=6em, text centered]
\tikzstyle{edgegui} = [draw, thick, color=cadmiumgreen!95, -latex', dashed]
\tikzstyle{edgepcs} = [draw, thick, color=alizarin!90, -latex', dashed]

\tikzstyle{edgeseq} = [draw, thick, color=cadmiumgreen!95, -latex']
\tikzstyle{edgesync} = [draw, thick, color=cornellred!95, -latex', dashed]

\tikzstyle{edgeasync} = [draw, thick, color=britishracinggreen!95, -latex', dashed]

\tikzset{
  startstop/.style={
  	rectangle, rounded corners,
  	minimum width=3cm, minimum height=1cm,
  	text centered, draw=black, fill=red!30	
  },
  process/.style={
    rectangle, 
    minimum width=3cm, minimum height=1cm, 
    text centered, draw=black, fill=orange!30	
  },
  decision/.style={
    diamond, 
    minimum width=3cm, minimum height=1cm, 
    text centered, draw=black, fill=green!30	
  },
  io/.style={
  	trapezium, 
	trapezium left angle=70, 
    trapezium right angle=110, 
    minimum width=1cm, 
    minimum height=0.5cm, 
    text centered, draw=black, fill=blue!30	
  },
  multidocument/.style={
  	draw,
  	tape bend top=none,
  	double copy shadow
  },
  manual input/.style={
  	shape=trapezium,
  	draw,
  	shape border rotate=90,
  	trapezium left angle=90,
  	trapezium right angle=80},
  cfg/.style={
  	draw,
  	rectangle,
  	rounded corners, 
  	minimum height=5mm, 
  	minimum width=18mm,
  	fill=white
  },
  callbackunit/.style={
  	rectangle, rounded corners,
  	minimum width=3cm, minimum height=0.6cm,
  	text centered, draw=black, fill=blanchedalmond	
  },	
  nodeno/.style={draw,circle,draw=cadetgrey,fill=cadetgrey,minimum size=3mm, inner sep=0pt},
  dummy/.style={rectangle},
  state2/.style={
  	draw=javapurple!95, 
  	fill=javapurple!95,
  	circle	
  },
  callbackcfg/.style={
  	callbacktext,
  	thick,
  	draw=black,
  	text width=8em,
  	minimum height=8em,
  	minimum width=12em,
  },
  callback/.style={
  	very thick,
  	draw=mikadoyellow,
  	top color=white,
  	bottom color=mikadoyellow!80!black!20, 
  },
  predicate/.style={
  	very thick,
  	draw=alizarin,
  	top color=white,
  	bottom color=alizarin!80!black!20, 
  },
  update/.style={
		very thick,
		draw=coolblack,
		top color=white,
		bottom color=coolblack!80!black!20, 
  },
  async/.style={
  		very thick,
  		draw=gray,
  		top color=white,
  		bottom color=gray!80!black!20, 
  },
  box/.style={
		rectangle,
		draw=black, 
		minimum height=6mm,minimum width=17mm, 
		fill=black!10, text badly centered, 
		font=\bfseries\footnotesize\sffamily
  },
  input/.style={
		box,
		fill=brightgreen!20,
		rounded corners,
  },
  inputback/.style={
	input,
	drop shadow,
  },
  process/.style={
		box,
		fill=gray!20,	
  },
  output/.style={
		box,
		fill=cinnabar!30,
  },
  outputback/.style={
		output,
		drop shadow,
  },
  iccfg/.style={
		box,
		diamond,
		fill=mikadoyellow!30,
  },
}

\newcommand{\splitatcommas}[1]{%
	\begingroup
	\begingroup\lccode`~=`, \lowercase{\endgroup
		\edef~{\mathchar\the\mathcode`, \penalty0 \noexpand\hspace{0pt plus 1em}}%
	}\mathcode`,="8000 #1%
	\endgroup
}

\newcommand{\shortauthors}{Dominguez Perez et al.}

\begin{document}
	\title{Testing Criteria for Mobile Apps \\Based on Callback Sequences}
	
	\author{\IEEEauthorblockN{1\textsuperscript{st} Danilo Dominguez Perez}
		\IEEEauthorblockA{\textit{Department of Computer Science} \\
			\textit{Iowa State University}\\
			Ames, IA \\
			danilo0@iastate.edu}
		\and
		\IEEEauthorblockN{2\textsuperscript{nd} Wei Le}
		\IEEEauthorblockA{\textit{Department of Computer Science} \\
			\textit{Iowa State University}\\
			Ames, IA \\
			danilo0@iastate.edu}
	}	
	
	\maketitle
	
	\begin{abstract}	
App quality has been shown to be the most important indicator of app adoption. To assure quality, developers mainly use testing to find bugs in app and apply {\it structural} and {\it GUI} test coverage criteria. However, mobile apps have more behaviors than the GUI actions, e.g. an app also handles events from sensors and executes long-running background tasks through Android API calls to \textit{Services} and \textit{AsyncTasks}. Our studies found that there are important app behaviors via callback interactions that should be covered in testing, as data sharing between callbacks is common and is the cause of many existing bugs. We design a family of test criteria based on callback sequences and use the {\it Callback Control Flow Automata (CCFA)} to measure the coverage for testing. Our experiments show that guiding by our criteria, testing can find more bugs and trigger bugs faster than the state-of-the-art tools.

\end{abstract}


	\maketitle
	
	\newcommand{\ee}{{\it event-event}\@\xspace}
	\newcommand{\eas}{{\it event-API sync}\@\xspace}
	\newcommand{\eaa}{{\it event-API async}\@\xspace}
	
	\newcommand{\fsm}{FSM\@\xspace}
	\section{Introduction}\label{sec:intro}
Since 2016, mobile devices have become the main medium to access the Internet \cite{mobilebrowsing}. The growth of the mobile ecosystem can also be seen in the millions of apps that are published~\cite{numberapps}. This growth creates a competition among apps in which the low quality apps get bad user reviews and end up being detrimental to their adoption \cite{appratings, appsremoved}. One of the main techniques that developers use to assure quality of the apps is testing.  Many tools have been developed to detect bugs in Android apps using automatic testing \cite{monkey, sapienz, stoat, dynodroid, a3e, guiripper, swifthand}. The majority of these tools focus on analyzing and testing \textit{Graphical User Interaces} (GUI) \cite{autotestingsurvey}. These approaches work on generating tests that maximize the coverage of a test criterion based on GUI event sequences \cite{coveragecriteriagui} and/or based on structural coverage such as statement and branch coverage. 


We believe that such criteria do not completely address the adequacy of test suites for mobile apps. Mobile apps, besides a rich GUI, also use other constructs to accomplish tasks. For example, background tasks executed through \textit{Services} or \textit{AsyncTasks} are not included in the GUI models. Moreover, the callbacks of these background tasks can interleave with GUI and other external event callbacks. Callbacks of other external events, such as GPS location updates or sensors, can also interleave with GUI callbacks. A coverage criterion based on GUI event sequences is not able to distinguish all the possible interleavings of these callbacks.  To test such code, we need to consider not just the execution behavior from the user through GUI, but also from the API calls to the framework and external components such as the camera and the GPS.

In this paper, we introduce white-box coverage criteria for testing Android apps based on the execution of callbacks. Specifically, we consider coverage criteria based on callback sequences, as they represent the behaviors that occur between different components of the mobile apps.



The challenge of developing such criteria is that the number of possible sequences of callbacks can be intractable. We thus need to evaluate what types of callbacks are important and what is the appropriate length of sequences to cover. We perform an empirical study to understand how values are propagated between callbacks and what types of callbacks more commonly share data. We then design coverage criteria to test the interactions of these callbacks. We also performed a bug study to find what types of callback interactions often lead to buggy behaviors. We thus should prioritize such callback interactions in testing. 

Based on our studies, we designed 3 callback coverage criteria for testing Android apps, namely \ee, \eas and \eaa.  The \ee criterion is designed to cover callback interactions between event handlers, including the handlers for GUI events and other external events such as GPS and sensors events. The \eas criterion aims to cover the sequence of an event handler and  the synchronous callbacks invoked in its API methods. The \eaa criterion is for exploring different concurrent behaviors via interleavings of an event handler and the asynchronous callbacks invoked in its API methods. 

To measure the coverage, we used a static representation called {\textit{Callback Control Flow Automata} (CCFA) \cite{DaniloTSE19}}. This model specifies all possible callback sequences in an app, including the callbacks invoked asynchronously and synchronously from external events and in API methods. We develop algorithms to statically compute the "ground truth" regarding which callback sequences in the app should be covered for each criterion. We instrument apps to generate the traces of callback sequences (the traces log the execution of entry and exit points of each callback). To obtain the coverage, we compare whether any ground truth callback sequences are included in the callback traces generated from testing. 

We implemented our coverage computation algorithms using Soot~\cite{vallee1999soot} on top of CCFA. We instrumented the apps using logcat~\cite{logcat} and Jacoco~\cite{jacoco} to collect traces during testing. We used 15 open source Android apps that these tools can handle. We tested these apps guided by our criteria and compared our results with the ones generated by {\it Monkey},  one of the best tools we can find for testing mobile apps~\cite{Choudhary:2015:ATI:2916135.2916273, Wang:2018:ESA:3238147.3240465}. Using the testing guided by our criteria, we found a total of 17 bugs, 3 more bugs than Monkey. Importantly, our testing is 7 times faster on average than Monkey to trigger the same set of bugs. For a total of 31 bugs reported by the two approaches, 30 bugs occurred with the increases of the coverage of our criteria.


In summary, this paper makes the following contributions: 
\begin{itemize}
	\item the empirical studies to show the importance of testing callback interactions ( \S III),
	\item the design of three test coverage criteria based on callback sequences ( \S IV),
	\item the approach of measuring the coverage criteria using a callback control flow graph, the CCFA, ( \S V) and 
	\item the evaluation that shows the importance of our coverage criteria based on real bugs found in the apps ( \S VI)
\end{itemize}

	\section{Motivation}\label{sec:motivation}
In this section, we present two real-world bugs to show the importance of guiding testing via desired callback sequences. 

\subsection{issue \#140~\cite{chanu140} in the app \textit{chanu}}
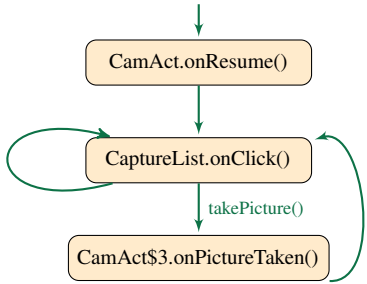
\begin{figure}
	\centering
	\begin{subfigure}[b]{0.86\columnwidth}
\begin{lstlisting}
class CamAct extends Activity {
  PictureCallback mP = new PictureCallback() {
    void onPictureTaken(byte[] d,Camera camera) {
      ...
+     mTaken = true;             
    }
  };
  void onResume() {
    cam = getCameraInstance();
    ...
    CaptureList l = new CaptureList();
    captureButton.setOnClickListener(l);    
  }
}
class CaptureList extends OnClickListener {
  void onClick(View v) {
+   if (mTaken) {
      cam.takePicture(null, null, mP);
+     mTaken = false; 
+   }
  }
]
\end{lstlisting}
		\caption{Issue \#140 in chanu \cite{chanu140}} \label{fig:chanu140code} 	
	\end{subfigure}
	~
	\begin{subfigure}[b]{\columnwidth}
		\centering
		\begin{tikzpicture}[->,>=stealth',shorten >=1pt,auto,node distance=2.4cm]
		
		\node[dummy]         (Dummy1)      at (0, 0.9)            {}; 
		\node[callbackunit] (OnResume)     at (0, 0)              {\footnotesize CamAct.onResume()};
		\node[callbackunit] (OnClick)      at (0, -1.3)           {\footnotesize CaptureList.onClick()};
		\node[callbackunit] (OnPicture)    at (0, -2.6)           {\footnotesize CamAct\$3.onPictureTaken()};
		\node[dummy]        (Dummy2)       at (2, -3.5)           {}; 
		\node[dummy]        (Dummy3)       at (-2, -3.5)            {}; 
		
		\draw[edgeseq]             (Dummy1)         -- node {}  (OnResume); 
		\draw[edgeseq]             (OnResume)       -- node {}  (OnClick);
		\draw[edgeseq]             (OnClick)        edge [loop left] node {}  (OnClick);
		\draw[edgeseq]             (OnClick)        -- node {\scriptsize takePicture()}  (OnPicture);
		\draw[edgeseq]             (OnPicture)      edge[bend right=100] node [left] {}  (OnClick);
		
		
		\end{tikzpicture}
		\caption{Sequences of callbacks executed in chanu}\label{fig:chanu140subseq}
	\end{subfigure}
	\caption{Code fragment and possible sequences of callbacks for chanu isssue \#140}~\label{fig:motivatingexamples}
\end{figure}

Here, we analyze issue \#140~\cite{chanu140} reported in the app \textit{chanu}. This bug is caused by an unexpected interleaving between a callback of an event handler and a callback invoked in the API method. In \figref{chanu140code}, we show three relevant callbacks that support the functionality of taking-a-picture in \textit{chanu}, including \inlinecode{onPictureTaken()}, \inlinecode{onResume()} and \inlinecode{onClick()}.


When taking a picture, the user clicks the GUI button, the system executes the callback \inlinecode{onClick()}, which then calls the API method \inlinecode{takePicture()} (at line~18). This method triggers an asynchronous task to capture an image and execute the callback \inlinecode{onPictureTaken()}.  \figref{chanu140subseq} shows the possible control flow of these callbacks. Along path \inlinecode{onClick()}  $\to$ \inlinecode{onClick()}, the user double-clicks a button, and along path \inlinecode{onClick()}  $\to$ \inlinecode{onPictureTaken()}   $\to$ \inlinecode{onClick()}, the user clicks the button, waits until the picture is taken, and then clicks the button again.

The bug is reported along \inlinecode{onClick()}  $\to$ \inlinecode{onClick()}. When the framework executes the callback \inlinecode{onClick()} for the second time before \inlinecode{onPictureTaken()}, the API method \inlinecode{takePicture()} crashes with a run-time exception. The root cause is that the camera is still busy responding to the first \inlinecode{onClick()}, and there is a race condition on a global flag in \inlinecode{takePicture()}. Whereas, along path \inlinecode{onClick()} $\to$ \inlinecode{onPictureTaken()}  $\to$ \inlinecode{onClick()},  \inlinecode{takePicture()} has finished \inlinecode{onPictureTaken()} for the first click and can proceed with the second click.  To fix this bug, the developer added a flag \inlinecode{mTaken} to avoid calling \inlinecode{takePicture()} while the camera is busy (see the fixes at lines 5, 17, 19 and 20 in \figref{chanu140code}).

Applying testing that aims to cover all statements or all callbacks, we can achieve 100\% coverage and stop testing after covering the path \inlinecode{onResume()}  $\to$  \inlinecode{onClick()} $\to$  \inlinecode{onPictureTaken()}. Such testing can miss the bug. Event-based testing~\cite{coveragecriteriagui} aims to cover the click event.  It does not distinguish the paths \inlinecode{onClick()}  $\to$ \inlinecode{onClick()} and  \inlinecode{onClick()}  $\to$ \inlinecode{onPictureTaken()}   $\to$ \inlinecode{onClick()}. We thus can also miss the bug.

\subsection{issue \#610~\cite{filedownloader610} in the app \textit{FileDownloader}}
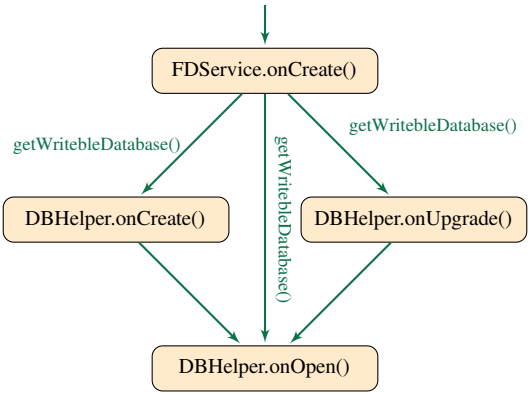
\begin{figure}
	\centering
	\begin{subfigure}[b]{0.86\columnwidth}
\begin{lstlisting}
class FDService extends Service {
  void onCreate() {
    ...
    DBHelper dbh = new DBHelper(this);
    db = dbh.getWritableDatabase();
  }
}
class DBHelper extends SQLiteOpenHelper {
  DBHelper(Context context) {
-   setWriteAheadLoggingEnabled(true);
  }
  void onOpen(SQLiteDatabase db) {
    setWriteAheadLoggingEnabled(true);
  }
  void onCreate(SQLiteDatabase db) {
    db.execSQL("CREATE TABLE IF NOT EXISTS" + ...);
  }
  void onUpgrade(SQLiteDatabase db, int ov, int nv) {
    if (oldVersion < 2) {
        String addColumn = "ALTER TABLE " ...;
        db.execSQL(addColumn);
    }
  }

\end{lstlisting}
		\caption{Issue \#610 in FileDownloader \cite{filedownloader610}} \label{fig:filedownloder610code} 	
	\end{subfigure}
	~
	\begin{subfigure}[b]{\columnwidth}
		\centering
		\begin{tikzpicture}[->,>=stealth',shorten >=1pt,auto,node distance=2.8cm]
		
		\node[dummy]         (Dummy1)      at (0, 0)            {}; 
		\node[callbackunit,align=center,] (OnCreate)     at (0, -1)  {\footnotesize FDService.onCreate()};
		\node[callbackunit] (OnCreate1)    [below left of=OnCreate]             {\footnotesize DBHelper.onCreate()};
		\node[callbackunit] (OnUpgrade)    [below right of=OnCreate]                {\footnotesize DBHelper.onUpgrade()};
		\node[callbackunit] (OnOpen)       [below right of=OnCreate1]           {\footnotesize DBHelper.onOpen()};
		%
		\draw[edgeseq]              (Dummy1)         -- node {}  (OnCreate); 
		\draw[edgeseq]             (OnCreate)       -- node [left] {{\scriptsize getWritebleDatabase()}}  (OnCreate1);
		\draw[edgeseq]             (OnCreate)       -- node {{\scriptsize getWritebleDatabase()}}  (OnUpgrade);
		\draw[edgeseq]             (OnCreate1)       -- node {}  (OnOpen);
		\draw[edgeseq]             (OnUpgrade)       -- node {}  (OnOpen);
		\draw[edgeseq]             (OnCreate)       -- node [sloped,anchor=south,auto=false] {{\scriptsize getWritebleDatabase()}} (OnOpen);
		
		\end{tikzpicture}
		\caption{Sequences of callbacks executed in FileDownloader}\label{fig:filedownloader610subseq}
	\end{subfigure}
	\caption{Code fragment and possible sequences of callbacks for FileDownloader isssue \#610}~\label{fig:motivatingexamples2}
\end{figure}

Here we show a bug that only can be triggered when we consider different callback sequences in an API method. In \figref{filedownloder610code}, we show a code snippet of a {\it service} that uses the SQLite APIs provided by the Android framework to access a database. When the API method \textit{getWritebleDatabase()} is called at line~5, it prepares the database \inlinecode{db} and returns \inlinecode{db}. There are a set of paths implemented in this API method, shown in \figref{filedownloader610subseq}. The path \inlinecode{FDService.onCreate()} $\to$  \inlinecode{DBHelper.onCreate()} represents the case where the app is being freshly installed.  The path \inlinecode{FDService.onCreate()} $\to$  \inlinecode{DBHelper.onUpgrade()} corresponds to the case where the app is updating an old version. The bug is located on the second path. In \inlinecode{FDService.onCreate()} at line~4, the constructor \inlinecode{DBHelper()} (see its implementation at line~9) is invoked. The function turns on "logging" functionality to maintain a copy of cache for the database. The bug is found along the path \inlinecode{FDService.onCreate()} $\to$  \inlinecode{DBHelper.onUpgrade()}. Here, \inlinecode{DBHelper.onUpgrade()} updates the database schema, but when using with \inlinecode{setWriteAheadLoggingEnabled()} (see line 10), the update is not visible immediately.


In this case, there are two different behaviors that could happen after \textit{getWritebleDatabase()}, depending on the environment in which the app runs. If the app is a fresh install, testing only covers \inlinecode{FDService.onCreate()} $\to$  \inlinecode{DBHelper.onCreate()}. To detect the bug, we need to exercise the path \inlinecode{FDService.onCreate()} $\to$  \inlinecode{DBHelper.onUpgrade()}, where the app updates a previously installed version. The event-based criteria are not able to find such bugs.




	\section{Which callback sequences we should test: An Empirical Study}\label{sec:studies}

In this section, we conduct studies to further investigate why it is important to test callback sequences. Specifically, in the first study, we study whether there can exist dataflow between callbacks and what types of callbacks share dataflow. We want to test such inter-callback paths, as any incorrect definition or use can only be exposed through executing the relevant callback sequences. In the second study, we explore what types of callback interactions are likely buggy and have reported bugs. We need to especially test such interactions of callbacks to avoid similar bugs in different apps. 



\subsection{What types of callbacks share dataflow?} 

We performed a def-use analysis for 55 apps randomly selected from the Google Play Market\footnote{\url{https://play.google.com}}. We used CCFA \cite{DaniloTSE19} as a callback control flow graph. We report that an {\it inter-callback dataflow} is found when there is a definition whose use is in a different callback. Our results show that all 55 apps contain inter-callback dataflow. Among all the def-use pairs we computed, 22.76\% are inter-callback. 

We performed further analysis to identify the type of callbacks in which def-use pairs are located. We found that 55\% of the inter-callback dataflow occurs between the callbacks that respond to external events, namely {\it event callbacks}. This is because the majority of the apps contain functionalities related to the GUI or the sensor components such as GPS or camera. The second most common (34\% of) inter-callback dataflow is located between synchronous callbacks invoked by the API method, namely {\it API\_SYNC callbacks}. This is because many API methods contain more than one synchronous callbacks, and it is common to share data within an API method. We also found def-use pairs between event callbacks and callbacks invoked in API methods, specially 7\% between event and API\_SYNC callbacks and 6\% between event and {\it API\_ASYNC} callbacks. API\_ASYNC are asynchronous callbacks invoked in API methods typically for responding to the {\it messages} from the API method. This interaction represents the scenario where the event handler invokes an API method and passes the data to the API method for handling the event. Our results are summarized in Table~\ref{tab:dataflow}.


\begin{table}[!ht]
	\centering
	\caption{The types of callback share dataflow. EV: event callback, AS: API\_SYNC callback, AA: API\_ASYNC callback}\label{tab:dataflow}
	\begin{tabular}{|c|c|c|c|} \hline
		EV-EV & AS-AS & EV-AS & EV-AA  \\ \hline \hline
		55\% & 34\% & 7\% & 6\% \\ \hline
	\end{tabular}
\end{table}

Importantly, our study also found that although the global dataflow can propagate through a maximum of 18 callbacks, 38\% global dataflow are related to consecutive callbacks along the paths in the callback control flow graph.

\subsection{What types of callback sequences can lead to bugs?} 

As a pilot study, we analyzed 1526 bugs from 6 apps that lead to crashes and found that 26 of such bugs are related to callback interactions; that is, we need to execute a sequence of callbacks to trigger the crash.  Furthermore, we found that all the 26 bugs are related to two callbacks, and 85\% of the bugs are related to the two consecutive callbacks. This result aligns with our findings from the first study and show that many of the cases, the data sharing occurs at the neighbor callbacks along the execution paths. 

In Table \ref{tab:bugstudy}, under {\it App}, we present the apps we studied. Under {\it Bugs}, we list the number of bugs we inspected. Under {\it Multi-C}, we give the number of bugs whose root causes are related to multiple callbacks. Under  {\it EV-EV},  {\it AS-AS},  {\it EV-AS}  and {\it EV-AA}, we show what types of callback interactions that lead to the bug.  Our results show that there are 6 bugs related to interactions of two event handlers, 11 bugs are related to synchronous calls in API methods, and 13 bugs are caused by not being able to handle all the interleavings correctly between the event handlers and the asynchronous callbacks in the API methods.

\begin{table}[!ht]
	\centering
	\caption{Bug study results}\label{tab:bugstudy}
	\resizebox{\columnwidth}{!}{
		\begin{tabular}{|l||c|c||c|c|c|c|} \hline
			App & Bugs & Multi-C & EV-EV & AS-AS & EV-AS & EV-AA  \\ \hline \hline
			ConnectBot & 9 & 5 & 1 & 2 & 0 & 2  \\ \hline
			FileDownloader & 12 & 2 & 0 & 1 & 1 & 0  \\ \hline
			AntennaPod & 133 & 5 & 1 & 1 & 0 & 4  \\ \hline
			cgeo & 273 & 4 & 2 & 1 & 0 & 3  \\ \hline
			Wordpress & 315 & 5 & 1 & 0  & 2 & 2  \\ \hline
			Ankidroid & 784 & 5 & 1 & 2 & 1 & 2  \\ \hline
			\textbf{Total} & \textbf{1526} & \textbf{26} & \textbf{6} &  \textbf{7} & \textbf{4} & \textbf{13}  \\ \hline
		\end{tabular}
	}
\end{table}

\subsection{Conclusions of the studies.} 
From the studies, we learned that although it is beneficial to test long callback sequences, the priority is to thoroughly cover neighbor callback interactions, starting with a length of two callbacks.

We also identified three important types of callback sequences to test. First, we want to test interactions between event handlers as these callbacks share data and they occupy the main behavior of apps. Second, synchronous callbacks invoked in the API methods are typically not targeted by any GUI testing tools; however, there are data sharing and bugs related to such callbacks. Finally, a main source of callback interaction bugs goes to race conditions caused by asynchronous invocations via event handlers and API methods. We need to sufficiently test the interleavings of these callbacks. 

\section{Defining Test Criteria Based on Callback Sequences}
Based on our findings, we designed a family of test criteria targeting the three important types of callback sequences. In this section, we first provide some background of CCFA~\cite{DaniloTSE19}. We then present our test criteria.



\subsection{Specifying Callback Sequences Using CCFA}
A CCFA is a representation based on the \textit{Extended Finite State Machine} (EFMS) \cite{1600197}. The goal is to specify all possible callback sequences identified from the app source code. There are 4 types of control flow between callbacks: 1) a callback $\mathit{B}$ is invoked synchronously after another callback $\mathit{A}$ is finished, 2) $\mathit{B}$ is invoked asynchronously after $\mathit{A}$, meaning that $\mathit{B}$ is put in the event queue after $\mathit{A}$, 3) during an execution of $\mathit{A}$, $\mathit{B}$ is invoked synchronously by an API call, and 4) during an execution of $\mathit{A}$, $\mathit{B}$ is invoked asynchronously by an API call, meaning that the API call puts $\mathit{B}$ in the event queue and the callback will be invoked eventually. 

In CCFAs, we use $\mathit{callbackname_{entry}}$ and $\mathit{callbackname_{exit}}$ as input symbols on the transitions. Each transition has a guard, specifying under which condition, the change of control between callbacks happen.  The callback can be triggered by an external event or invoked by an API method.

\begin{figure}
	\centering
	\begin{subfigure}[b]{0.75\columnwidth}
		\centering
		\resizebox{0.95\columnwidth}{!}{
\begin{lstlisting}
class A extends Activity {
  Button b1;
  void onCreate(Bundle b) {
    b1 = (Button) findViewById(...);
	b1.setOnClickListener(new CList());
  }
  void onStart() {
    ...
    if (*) {
      L l = new L();
      LoaderManager lm = loaderManager();
      lm.initLoader(0, null, l); 
   	}
  }
  void onStop() { ... }
}
class L implements LoaderCallbacks {
  Loader onCreateLoader(int i, 
  						Bundle b) {
    ...
  } 
}
class CList extends OnClickListener {
  void onClick(View v) { ... }
}
\end{lstlisting}
		}
		\caption{Source code of a simple Android app\label{fig:examplecode}}		
	\end{subfigure}
	\begin{subfigure}[b]{\columnwidth}
		\centering
		\resizebox{\columnwidth}{!}{
			\begin{tikzpicture}[->,>=stealth',shorten >=1pt,auto,node distance=3.45cm,
			semithick]
			
			\node[initial,state, accepting] (Q1)                    {$q_1$};
			\node[state]         (Q2)  [above right=of Q1, xshift=-0.6cm] {$q_2$};
			\node[state]         (Q3)  [right of=Q2] {$q_3$};
			\node[state]         (Q4)  [right of=Q3] {$q_4$};
			\node[state]         (Q5)  [below of=Q3,xshift=0.7cm] {$q_5$};
			\node[state]         (Q6)  [right of=Q4, xshift=-1cm] {$q_6$};
			\node[state]         (Q8)  [below of=Q6] {$q_7$};
			\node[state]         (Q9)  [below of=Q8] {$q_{8}$};
			\node[state]         (Q10) [left of=Q9,  xshift=1cm] {$q_{9}$};
			\node[state]         (Q11) [below of=Q2, xshift=-0.7cm] {$q_{10}$};
			\node[state]         (Q12) [below of=Q5] {$q_{11}$};
			
			\path
			(Q1) edge[edgeasync]         node[align=center] {($\mathit{A.onCreate_{entry}}$, \\  $\mathit{evt = launch}$)} (Q2)
			(Q2) edge          node[align=center] {($\mathit{A.onCreate_{exit}}$, \\ $\mathit{true}$)} (Q3)
			(Q3) edge          node[align=center] {($\mathit{A.onStart_{entry}}$, \\ $\mathit{true}$)} (Q4)
			(Q4) edge         node[align=right, near end] {\footnotesize ($\mathit{A.onStart_{exit}}$, $\mathit{true}$)} (Q5)
			edge          node[align=center] {$\epsilon$} (Q6)
			(Q6) edge[swap]          node[align=center] { ($\mathit{L.onCreateLoader_{entry}}$, \\  $\mathit{cs=lm.initLoader()}$)} (Q8)
			
			(Q8) edge[swap]          node[align=center] {($\mathit{L.onCreateLoader_{exit}}$, \\ $\mathit{true}$)} (Q9)
			(Q9) edge          node[align=center] {$\epsilon$} (Q10)    
			(Q10) edge         node[align=left, right, very near end] {($\mathit{A.onStart_{exit}}$, $\mathit{true}$)} (Q5)
			
			(Q5) edge[edgeasync, bend left]         node[align=center,swap] {(\footnotesize $\mathit{CList.onClick_{entry}}$, \\ \footnotesize $\mathit{evt = click_{b1}}$)} (Q11)
			(Q11) edge[bend left]         node[align=center] {($\mathit{CList.onClick_{exit}}$, \\ $\mathit{true}$)} (Q5)
			(Q5) edge[edgeasync]     node[align=right, left] {($\mathit{A.onStop_{entry}}$, \\ $\mathit{evt = back\_button}$)} (Q12)
			(Q12) edge[bend left]     node[align=center] {($\mathit{A.onStop_{exit}}$, \\ $\mathit{true}$)} (Q1)
			;
			\end{tikzpicture}
		}
		\caption{A CCFA for simple app}\label{fig:ccfa}
	\end{subfigure}
	\caption{An example app and its CCFA}~\label{fig:ex}
\end{figure}
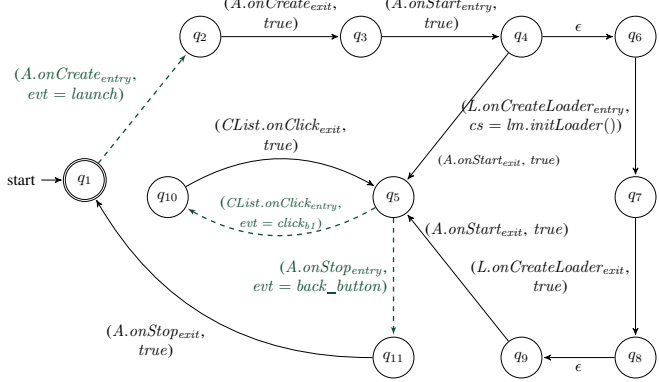

\figref{ex} shows a simple Android app and its CCFA. The app has five callbacks.  The paths on the CCFA provides the possible sequences of these 5 callbacks. At the initial state $q1$, \inlinecode{onCreate()} from \inlinecode{class A} is invoked asynchronously when the event {\it launch} is triggered, noted by the transition $A.onCreate_{entry}, evt=launch$. This callback is followed synchronously by \inlinecode{onStart()} from \inlinecode{class A}. During the execution of \inlinecode{onStart()} , the API call \inlinecode{lm.initLoader(0, null, 1)} is invoked, which calls \inlinecode{onCreateLoader()} from \inlinecode{class L} synchronously (see transitions from $q4\rightarrow q_9$). At $q_5$, \inlinecode{onClick()} from \inlinecode{class CList} can be invoked asynchronously any number of times until  \inlinecode{onStop()} from \inlinecode{class A} is invoked.




\subsection{Testing Criteria Based on Callback Sequences}


\subsubsection{Preliminaries} We first formally introduce the concepts of {\it callback sequence} and {\it trace}. 

\begin{mydef}[Callback Sequence]
	Let $C = \{c1, c2, ..., cn\}$ be the callbacks implemented for an app $P$. The input symbols of the CCFA for $P$ are defined by the set $I = \{c1_{entry}, c1_{exit}, c2_{entry}, c2_{exit}, ..., cn_{entry}, cn_{exit}\}$. A {\it callback sequence} is $\langle s1 \to s2 \to ...\to sm \rangle$ where $si \in I$ for $1 \leq i \leq m$. The {\it length} of a callback sequence is the number of callbacks, or, in another words, the number of the input symbol $c_{entry}$ in the sequence.
\end{mydef}

\begin{mydef}[Trace]
	By instrumenting the entry and exit of callbacks, during testing, we generate a callback sequence for each execution, which we call {\it trace}.
\end{mydef}

\subsubsection{The \ee Criterion}
 This criterion is designed to cover callback interactions related to event callbacks. The events can be GUI events or other external events, e.g. the ones generated by the sensors. 

\begin{mydef}[the \ee criterion]\label{def:c1}
	Let $C=\{c1, c2, ..., cn\}$  be the callbacks implemented for an app $P$.  Let $E \subseteq C$ be the set of callbacks that handle the events in $P$. Let $I_E$ be the set of input symbols (the entry and exit points) for any callback $c \in E$. The \ee coverage criterion is satisfied if and only if for all callback sequences of length 2,  $\langle s1 \to s2 \to s3\rangle$, where $s1,s2,s3 \in I_{E}$, generated from the CCFA of $P$, there exists a trace $T$ that contains $S$. 
\end{mydef}

\begin{figure}
	\centering
	\begin{subfigure}[b]{\columnwidth}
		\centering
		\resizebox{0.82\columnwidth}{!}{
\begin{lstlisting}
class A extends Activity {
  Button b1;
  Button b2;
  LocList l;
  void onCreate(Bundle b) {
    ..
    b1.setOnClickListener(new Button1());
    b2.setOnClickListener(new Button2());
    lm.requestLocationUpdates(.., new LocList()); 
  }  
}
class LocList implements LocationListener {
  void onLocationUpdate(...) {
    ...
  } 
}
class Button1 extends OnClickListener {
  void onClick(View v) { ... }
}
class Button2 extends OnClickListener {
  void onClick(View v) { ... }
}
\end{lstlisting}
		}
		\caption{a simple Android app that handles the GPS event\label{fig:example3}}       
	\end{subfigure}
	\begin{subfigure}[b]{\columnwidth}
		\centering
		\resizebox{0.82\columnwidth}{!}{
			\begin{tikzpicture}[->,>=stealth',shorten >=1pt,auto,node distance=2.3cm,
			semithick]
			
			\node[initial,state] (Q1)                    {$q_1$};
			\node[state]         (Q2)  [below of=Q1] {$q_2$};
			\node[state]         (Q3)  [below of=Q2] {$q_3$};
			\node[state]         (Q4)  [left of=Q3, xshift=-0.9cm] {$q_4$};
			\node[state]         (Q5)  [right of=Q3, xshift=0.9cm] {$q_5$};
			\node[state]         (Q6)  [below of=Q3,yshift=-0.9cm] {$q_6$};
			
			\path
			(Q1) edge         node[align=center] {(\footnotesize $\mathit{A.onCreate_{entry}}$, $\mathit{evt = launch}$)} (Q2)
			(Q2) edge          node[near start] {(\footnotesize $\mathit{A.onCreate_{exit}}$, $\mathit{true}$)} (Q3)
			(Q3) edge[edgeasync, bend right,swap]          node[align=center] {(\footnotesize $\mathit{Button1.onClick_{entry}}$, \\ \footnotesize $\mathit{evt = click_{b1}}$)} (Q4)
			(Q4) edge[bend right, swap]          node[align=center] {(\footnotesize $\mathit{Button1.onClick_{exit}}$, \\ \footnotesize $\mathit{true}$)} (Q3)
			
			(Q3) edge[edgeasync, bend left]          node[align=center] {(\footnotesize $\mathit{Button2.onClick_{entry}}$, \\ \footnotesize $\mathit{evt = click_{b2}}$)} (Q5)
			(Q5) edge[bend left]          node[align=center] {(\footnotesize $\mathit{Button2.onClick_{exit}}$, \\ \footnotesize $\mathit{true}$)} (Q3)
			
			(Q3) edge[edgeasync, bend right, swap]          node[align=center,near end] {(\footnotesize $\mathit{LocList.onLocationUpdate_{entry}}$, \\ \footnotesize $\mathit{evt = location}$)} (Q6)
			(Q6) edge[bend right, swap]          node[align=center,near start] {(\footnotesize $\mathit{LocList.onLocationUpdate_{exit}}$, \\ \footnotesize $\mathit{true}$)} (Q3)
			;
			\end{tikzpicture}
		}
		\caption{A CCFA for the app}\label{fig:ccfaexample3}
	\end{subfigure}
	\caption{An app with Location Callbacks}\label{fig:applocation}
\end{figure}
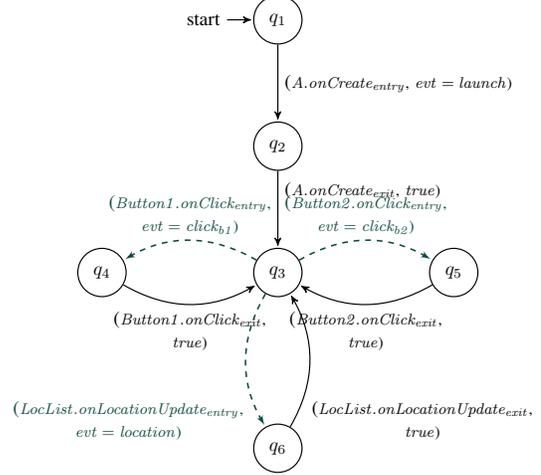

The behaviors required to be covered by the \ee criterion are close to the behaviors covered by GUI testing tools~\cite{guiripper}. The difference is that the \ee criterion also requires to exercise the interactions between callbacks invoked in other types of external events besides GUI. As an example, in Figure~\ref{fig:example3}, the app registers the listener callbacks for two buttons at lines~7 and 8, and a callback to receive GPS location updates at line~9. Figure \ref{fig:ccfaexample3} shows the CCFA of the app. Given the \ee criterion, we generate the following callback sequences starting \inlinecode{A.onCreate()}:

\footnotesize{
\begin{enumerate}
	\item $\mathit{A.onCreate_{entry}} \to \mathit{A.onCreate_{exit}} \to$  $\mathit{Button1.onClick_{entry}}$
	\item $\mathit{A.onCreate_{entry}} \to \mathit{A.onCreate_{exit}} \to$  $\mathit{Button2.onClick_{entry}}$
	\item $\mathit{A.onCreate_{entry}} \to \mathit{A.onCreate_{exit}} \to$ \\ $\mathit{LocList.onLocationUpdate_{entry}}$
\end{enumerate}
}
\normalsize
Using a similar approach, we can generate the callback sequence centered on \inlinecode{Button1.onClick} (including 2 sequences), \inlinecode{Button2.onClick} (2 sequences) and \inlinecode{LocList.onLocationUpdate} (2 sequences). That is, there are a total of 9 (3+2+2+2) callback sequences we should cover under the \ee criterion, including the cases where the location event happens before or after a GUI callback. Whereas, any GUI based coverage criterion does not exercise such behaviors. 

\subsubsection{The \eas Criterion} The second criterion \eas focuses to trigger callback interactions related to synchronous callbacks implemented in the API methods. Specifically, there are two types of sequences are of interest: 1)  a synchronous callback in the API method with the caller of the API method which is an event callback, and 2) two synchronous callbacks in the same API method. 


\begin{mydef}[the \eas criterion]
	Let $Z =\{c1, c2, ..., cn\}$ and $Z \subseteq C$ be the set of callbacks of type {\it API\_SYNC} in an app $P$, and $I_{Z}$ be the set of input symbols  (the entry and exit points)  for any callback $c \in Z$ in CCFA. Let $E \subseteq C$ be the set of callbacks for handling events in an app $P$, and $I_{E}$ be the set of input symbols for any callback $c \in E$. The \eas criterion is satisfied if and only if for all the callback sequences of length 2, $\langle s1 \to s2 \rangle$, where $s1 \in I_E$ or $s1\in I_{Z}$ and $s2 \in I_{Z}$, generated from the CCFA of $P$,  there exists a trace $T$ that contains $S$. 	
\end{mydef}


This test criterion requires to trigger not only the API method but also the synchronous call(s) in the API method. For example, in Figure~\ref{fig:ccfa}, we should cover $\mathit{A.onStart_{entry}} \to \mathit{L.onCreateLoader_{entry}}$ along the transitions from $q_3, q_4, q_6$ and $q_7$. Sometimes, there are different callback sequences in the API method, dependent on the state of the system when the API call is invoked. This criterion requires to test all possible callback sequences. As an example, in Figures \ref{fig:filedownloder610code} and \ref{fig:filedownloader610subseq}, the \eas criterion requires to cover all the following callback sequences from  \inlinecode{FDService.onCreate()}:

\footnotesize{
\begin{enumerate}
	\item $\mathit{FDService.onCreate_{entry}} \to \mathit{DBHelper.onCreate_{entry}}$
	\item $\mathit{FDService.onCreate_{entry}} \to \mathit{DBHelper.onUpgrade_{entry}}$
	\item $\mathit{FDService.onCreate_{entry}} \to \mathit{DBHelper.onOpen_{entry}}$
\end{enumerate}
}
\normalsize

To satisfy the criterion, the test need to setup different running states of an app: (1) the app is being installed, (2) the app is being updated from an older version previously installed on the phone, and (3) the app is running. 




\subsubsection{The \eaa criterion} The API methods can make asynchronous calls such as \inlinecode{Handler.sendMessage} and \inlinecode{Context.startService}. These methods will invoke the asynchronous callbacks. The past studies ~\cite{raceandroidmaiya, raceandroidhsiao, racebielik} as well as our own bug study presented in Section \ref{sec:studies} all indicate that asynchronous callbacks invoked in the API calls can lead to concurrency bugs. Such bugs typically involve two event calls and one API\_ASYNC.  For example, in Figures \ref{fig:chanu140code} and \ref{fig:chanu140subseq},  there are two possible interleavings between the callbacks of  \inlinecode{CaptureList.onClick()} and \inlinecode{CamAct\$3.onPictureTaken}: 

\footnotesize{
\begin{enumerate}
	\item $\mathit{CaptureList.onClick_{entry}}\to$  	
	 $\mathit{CaptureList.onClick_{exit}} \to$ \\
	 $\mathit{CamAct\$3.onPictureTaken_{entry}}\to $  \\
	 $\mathit{CamAct\$3.onPictureTaken_{exit}}\to $ 
	  $\mathit{CaptureList.onClick_{entry}}$
	\item $\mathit{CaptureList.onClick_{entry}} \to$ 
		  $\mathit{CaptureList.onClick_{exit}} \to$ \\
		  $\mathit{CaptureList.onClick_{entry}} \to$  
		  $\mathit{CaptureList.onClick_{exit}}\to $   \\
		  $\mathit{CamAct\$3.onPictureTaken_{entry}}$
\end{enumerate}
}
\normalsize


In the first case, there are no other tasks in the event queue when the task of \inlinecode{CamAct\$3.onPictureTaken()} was posted by \inlinecode{takePicture()}. As a result, \inlinecode{CamAct\$3.onPictureTaken()} is executed immediately after \inlinecode{CaptureList.onClick()}. In the second case, the second \inlinecode{CaptureList.onClick()} follows right after the first \inlinecode{CaptureList.onClick()} before \inlinecode{takePicture()} posts the task of \inlinecode{CamAct\$3.onPictureTaken()}. It is hard to handle all the interleavings correctly during implementation. Thus we should test the interleavings to help expose the bugs.

\begin{mydef}[The \eaa criterion]
	Let $Y =\{c1, c2, ..., cn\}$ and $Y \subseteq C$ be the set of API\_ASYNC callbacks. Let $I_{Y}$ be the set of input symbols (the entry and exit points) for any callback $c \in Y$. The \eaa criterion is satisfied if and only if for every callback sequence $S$ of length 3 computed using $f(c)$ ($c \in Y$) from the CCFA, there exists a trace $T$ that contains $S$. 	 
	
	 
\end{mydef}




Here is how $f(c)$ is computed: we first find $c\in Y$, and then we traverse the CCFA to find its caller, $e$. We identify any successor of $e$ on CCFA, $s$, which is also an event callback. $s$ and $c$ potentially run concurrently. We then list all the interleavings involving $e$, $s$ and $c$ on CCFA. In this paper, we focus on the interleaving between two event callbacks and one API\_ASYNC, as most of the bugs we found are caused by such interleavings. 

It should be noted that the \ee criterion also tests the sequences of two event callbacks. However, the \eaa criterion takes a step further to enforce all the interleaving between two event callbacks and an API\_ASYNC callback. The \ee criterion only requires to cover two events without enforcing any behavior related to API\_ASYNC.


	\section{Measuring Test Coverage}~\label{coverage}
In this section, we present our methodologies of calculating the coverage of three criteria in testing.

\begin{figure}
	\centering
	\includegraphics[width=\columnwidth]{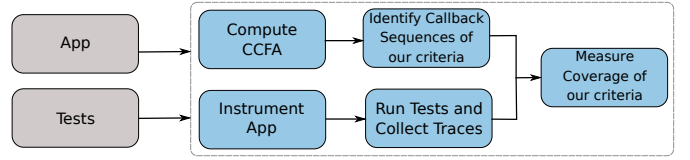}
	\caption{the Framework for Measuring Coverage}\label{fig:architecture}
\end{figure}

\subsection{The Framework for Measuring Coverage of Three Criteria}~\label{trace}
As shown in Figure~\ref{fig:architecture}, given an Android app (in the {\it APK} format) and its tests, we first apply a static analysis described in \cite{DaniloTSE19} to generate a CCFA from the APK file. Using the CCFA, we apply different traversal algorithms (described later in this section) to generate callback sequences for the three coverage criteria respectively. These are the callback interactions an ideal test set should cover for the app. 

We also analyze the CCFA to identify whether a callback is an event callback, API\_SYNC or API\_ASYNC. On the CCFA, the transitions of an event callback are labeled with its triggering event, the transitions labeled with the message represent the API\_ASYNC (these calls are triggered by the messages in the Android framework), and finally for any callbacks invoked in the API method (labeled with the API call site between the two $\epsilon$ edges on the CCFA), they are the API\_SYNC if not already marked as API\_ASYNC. This mapping between the callbacks and their types will be used for analyzing the trace to determine what callback interactions a test actually covers.

To collect the trace, we develop a tool that takes an APK file as an input. It identifies the entry and the exit of all the callbacks in the app and inserts the instrumentation. Each line in the trace prints a tuple $(t, s, k)$, where $t$ is the time when the instrumented statement was executed, $s$ is the signature of the callback and $k$ specifies whether it is an entry or exit point. As an example, a test that launches the app and then executes a \textit{click back-button} event in Figure \ref{fig:examplecode} generates the following trace, $T_{\ref{fig:examplecode}}$: 

\footnotesize{
\begin{lstlisting}
	(t1, A.onCreate(), ENTRY)
	(t2, A.onCreate(), EXIT)
	(t3, A.onStart(), ENTRY)
	(t4, L.onCreateLoader(), ENTRY)
	(t5, L.onCreateLoader(), EXIT)
	(t6, A.onStart(), EXIT)
	(t7, A.onStop(), ENTRY)
	(t8, A.onStop(), EXIT)
\end{lstlisting}
}
\normalsize





Given the traces, in the following, we show how to calculate the coverage using CCFA for each criterion.

\subsection{Measuring Coverage for the {\it Event-Event} Criterion}\label{sec:measurement}
Given an app, we first identify the "ground truth", that is, what callback sequences a test should cover for the \ee criterion. We traverse the CCFA and find all the two consecutive event callbacks (we name this set $S_{ee}$) on the CCFA. To do so, we first modify the CCFA and remove all the transitions between the $\epsilon$ transitions. The $\epsilon$ transitions mark the beginning and the end of the API calls. Thus this step removes all the API\_SYNC and API\_ASYNC callbacks in the CCFA, so we can consider only event callbacks which are needed for this criterion. Using the modified CCFA, we start at every transition that represents the entry point of a callback and generate sequences of two consecutive callbacks. For loops, we traverse the loop once. 

As an example, in Figure \ref{fig:ccfa}, we traverse the transitions $q_1$ to $q_4$ and generate the sequence $A.onCreate_{entry} \to A.onCreate_{exit} \to A.onStart_{entry}$.  Following the  $\epsilon$ transitions starting at $q_4$ and $q_8$, we remove the callbacks in the API methods located between $q_6$ and $q_9$. As a result, we obtain two sequences $A.onStart_{entry} \to A.onStart_{exit} \to CList.onClick_{entry}$ and $A.onStart_{entry} \to A.onStart_{exit} \to A.onStop_{entry}$ along $q_3 \to q_5 \to q_{10}$ and $q_3 \to q_5 \to q_{11}$ respectively.

To calculate the coverage, we compare the sequences computed from the CCFA, $S_{ee}$, with the traces. We first filter the trace to contain just event callbacks using the mapping of callbacks and their types pre-computed from the CCFA, resulting in the trace $T_{E} \subseteq T$. We then check whether a sequence $s \in S_{ee}$ is included by $T_{E}$.  For instance, before filtering, the trace $T_{\ref{fig:examplecode}}$ listed at the end of Section~\ref{trace} does not contain the required sequence $A.onStart_{entry} \to A.onStart_{exit} \to A.onStop_{entry}$. But after filtering, we generate the following sequence, which indicates that the required sequence is covered.

\footnotesize{
\begin{lstlisting}
(t1, A.onCreate(), ENTRY)
(t2, A.onCreate(), EXIT)
(t3, A.onStart(), ENTRY)
(t6, A.onStart(), EXIT)
(t7, A.onStop(), ENTRY)
(t8, A.onStop(), EXIT)
\end{lstlisting}
}
\normalsize


Let $C_{ee} \subseteq S_{ee}$ be the set of sequences actually covered by $T_{E}$. The \ee coverage is computed by $|C_{ee}|/|S_{ee}|$. Note that when there are multiple traces generated in testing, $C_{ee}$ will include the sequences covered by all the traces.


\subsection{Measuring Coverage for the Event-API Sync Criterion}
This criterion focuses on testing the synchronous callbacks in the API methods. Thus our first step is to remove all the asynchronous callbacks invoked in the API method (the transitions labeled with the message guard on CCFA). We then traverse the modified CCFA. When the traversal reaches the transition labeled with the entry point of an API\_SYNC, we find its predecessor callback (its caller). For example, in Figure \ref{fig:ccfa}, when we reach the transition $q_6$ to $q_7$, we perform a backward traversal to  identify the entry point of its caller $A.onStart_{entry}$. Thus, for the transitions $q_3 \to q_4 \to q_6 \to q_7$, we generate the callback sequence $A.onStart_{entry} \to L.onCreateLoader_{entry}$. Using a similar way, the traversal visits all the synchronous callbacks in the API methods and gets their predecessors to form the sequences.  


To calculate the coverage given the trace $T$, we check if any of the required sequence computed above, $s \in S_{eas}$, is covered by $T$. Note that in the trace, the two callbacks of API\_SYNC should always occur consecutively if they are consecutive on the CCFA, so we do not need to filter out any callbacks in the trace before checking. Let $C_{eas} \subseteq S_{eas}$ be the set of sequences covered by $T$. The coverage of the \eas criterion is $|C_{eas}|/|S_{eas}|$.  

\subsection{Measuring Coverage for the Event-API Async Criterion}
\begin{figure}
	\centering
	\resizebox{0.85\columnwidth}{!}{
		\begin{tikzpicture}[->,>=stealth',shorten >=1pt,auto,node distance=2.5cm,
		semithick]
		
		\node[initial,state] (Q1)                    {$q_1$};
		\node[state]         (Q2)  [below of=Q1] {$q_2$};
		\node[state]         (Q3)  [below of=Q2] {$q_3$};
		\node[state]         (Q4)  [left of=Q3, xshift=-1.2cm, yshift=0.7cm] {$q_4$};
		\node[state]         (Q5)  [left of=Q4] {$q_5$};
		\node[state]         (Q6)  [below of=Q5] {$q_6$};
		\node[state]         (Q7)  [below right of=Q6] {$q_7$};
		\node[state]         (Q8)  [right of=Q7, xshift=1cm] {$q_8$};
		
		\node[state]         (Q9)  [right of=Q3, xshift=1.4cm] {$q_9$};

		\path
		(Q1) edge[edgeasync]         node[align=center] {(\footnotesize $\mathit{CamAct.onResume_{entry}}$, $\mathit{evt = launch}$)} (Q2)
		(Q2) edge          node[near start] {(\footnotesize $\mathit{CamAct.onResume_{exit}}$, $\mathit{true}$)} (Q3)
		(Q3) edge[edgeasync, bend right,swap]          node[align=center, pos=0.65] {(\footnotesize $\mathit{CaptureList.onClick_{entry}}$, \\ \footnotesize $\mathit{evt = click_{capture}}$)} (Q4)
		(Q4) edge[swap]          node[align=center] {$\epsilon$} (Q5)
		(Q5) edge[edgeasync, bend right]          node[align=center] {(\footnotesize $\mathit{CamAct\$3.onPictureTaken_{entry}}$, \\ \footnotesize $\mathit{msg = cam.takePicture()}$)} (Q6)
		(Q6) edge[bend right]          node[align=center] {(\footnotesize $\mathit{CamAct\$3.onPictureTaken_{exit}}$, \\ \footnotesize $\mathit{true}$)} (Q7)
		(Q7) edge[swap]          node[align=center] {$\epsilon$} (Q8)
		(Q8) edge[bend right,swap]          node[align=center] {(\footnotesize $\mathit{CaptureList.onClick_{exit}}$, \\ \footnotesize $\mathit{true}$)} (Q3)
		
		(Q3) edge[edgeasync, bend left]          node[align=center] {(\footnotesize $\mathit{CaptureList.onTab_{entry}}$, \\ \footnotesize $\mathit{evt = tab_{tab1}}$)} (Q9)
		
		(Q9) edge[bend left]          node[align=center] {(\footnotesize $\mathit{CaptureList.onTab_{exit}}$, \\ \footnotesize $true$)} (Q3)
		;
		\end{tikzpicture}
	}
	
	\caption{The CCFA for Figure \ref{fig:chanu140code}}\label{fig:ccfachanu}
\end{figure}
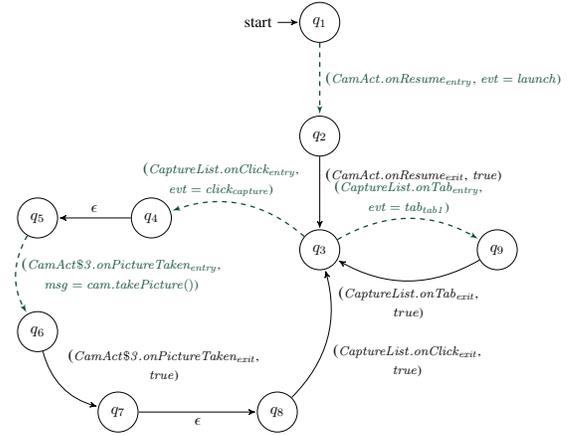

This criterion requires to test different interleavings involving the API\_ASYNC callbacks. To compute all the required callback sequences $S_{eaa}$, we first traverse the CCFA and find the entry points of the API\_ASYNCs. For example, in Figure~\ref{fig:ccfachanu} (the code is given in Figure \ref{fig:chanu140code}), we first identify the callback, \inlinecode{onPictureTaken()} indicated by the transition $q_5\to q_6$ (we call this transition $t$) . Our next step is to find the caller of such callback. This goal is achieved by traversing the CCFA backwards until we find an asynchronous transition. For the transition $t$, we identify $\mathit{CaptureList.onClick_{entry}}$ from the transition $q_3\to q_4$ (we call this transition $t_{caller}$). 

To find the asynchronous callbacks that potentially have a race condition with the callback at $t$, we start traversing the CCFA at $t_{caller}$, and find all of its possible asynchronous callback successors. These are the callbacks that can follow $t_{caller}$, and dependent on the timing, they are possibly executed before $t$ or after $t$, creating different interleavings among these callbacks. For example, given $t_{caller}$ found at $q_5 \to q_6$ in Figure \ref{fig:ccfachanu}, we find the next reachable asynchronous callbacks $\mathit{CaptureList.onClick}$ and $\mathit{CaptureList.onTab}$ from the transitions $q_3 \to q_4$ and $q_3 \to q_9$ respectively. Based on which we can generate the following four callback sequences:

 \footnotesize{
\begin{enumerate}
	\item $\mathit{CaptureList.onClick_{entry}}\to$  	
	$\mathit{CaptureList.onClick_{exit}} \to$ \\ 
	$\mathit{CamAct\$3.onPictureTaken_{entry}}\to $  \\
	$\mathit{CamAct\$3.onPictureTaken_{exit}}\to $  
	$\mathit{CaptureList.onClick_{entry}}$
	\item $\mathit{CaptureList.onClick_{entry}} \to$ 
	$\mathit{CaptureList.onClick_{exit}} \to$ \\
	$\mathit{CaptureList.onClick_{entry}} \to$  
	$\mathit{CaptureList.onClick_{exit}}\to $  \\ 
	$\mathit{CamAct\$3.onPictureTaken_{entry}}$
	\item $\mathit{CaptureList.onClick_{entry}}\to$  
	$\mathit{CaptureList.onClick_{exit}} \to$ \\ 
	$\mathit{CamAct\$3.onPictureTaken_{entry}}\to $ \\ 
	$\mathit{CamAct\$3.onPictureTaken_{exit}}\to $ 
	$\mathit{CaptureList.onTab_{entry}}$
	\item $\mathit{CaptureList.onClick_{entry}} \to$ 
	$\mathit{CaptureList.onClick_{exit}} \to$ \\
	$\mathit{CaptureList.onTab_{entry}} \to$  
	$\mathit{CaptureList.onTab_{exit}}\to $  \\ 
	$\mathit{CamAct\$3.onPictureTaken_{entry}}$
\end{enumerate}
}
\normalsize

Note that we apply these steps for each API\_ASYNC in the CCFA. For example, if there is another API\_ASYNC $x$ that is invoked after 	$\mathit{CamAct\$3.onPictureTaken}$ in the API method $cam.takePicture$, we generate all the interleavings among the callbacks $x$,  $\mathit{CaptureList.onClick}$ and $\mathit{CaptureList.onTab}$. Due to the timing issue,  $\mathit{CaptureList.onClick}$ and $\mathit{CaptureList.onTab}$ may be put in the event queue before or after $x$ is put in the queue.

To calculate the coverage for this criterion, we consider traces that involve API\_ASYNC and event callbacks. Similar to measuring the \ee criterion, for a trace $T$, we perform filtering and exclude all API\_SYNCs in the trace, resulting in $T_{eaa}  \subseteq T$. We say a sequence $s \in S_{eaa}$ is covered if $T_{eaa}$ contains $s$. Let $C_{eaa}$ be the set of sequences covered by $T_{eaa}$. The coverage of the \eaa criterion is computed by $|C_{eaa}|/|S_{eaa}|$.

	\section{Evaluation}
The goal of our evaluation is to empirically show that our coverage criteria can help testing quickly find certain types of bugs that other testing tools and criteria are difficult, slow and sometimes impossible to trigger. 




\subsection{Implementation and Experimental Setup}
We implemented the algorithms~\cite{DaniloTSE19} to generate CCFAs for Android apps using Soot \cite{vallee1999soot}. We also built a tool to instrument Android apps to collect traces using Soot. When running the instrumented apps, we used \textit{logcat}~\cite{logcat} to collect callback traces from the phone logs. To calculate the coverage of our criteria, we implemented the techniques described in Section~\ref{coverage}. We collected statement coverage by instrumenting apps using Jacoco \cite{jacoco}, and we calculated GUI coverage by identifying two consecutive GUI events on CCFAs and determining if they are in the trace.   


We performed experiments on 15 apps from the open source repository F-droid \cite{fdroid}, shown in Table \ref{tab:benchmarks}. We had planned a larger scale study. However, we faced a set of constraints: (1) the apps need to work with CCFA which are dependent on the tools of Soot, Gator and PCS~\cite{DaniloTSE19}, (2) the source code of the apps should be available so we can perform instrumentation using logcat and Jacoco, (3) we used apps that do not require special inputs from the user such as a username and a password, as done in other studies~\cite{sapienz}, and (4) it takes significant amount of time to manually inspect the bugs found to ensure they are valid bugs and also to inspect the callback sequences we generated to ensure they indeed confirm to our test criteria. We have included as many apps as we can in the experiments given the resources we had. These apps cover more than 10 categories with the largest size of 45.8 k lines of code.  We believe that they form  representative samples to reach valid conclusions. 



\begin{table}
	\centering
	\caption{The benchmark}\label{tab:benchmarks}
	\begin{tabular}{|l|l|r|r|} \hline
		App & Category & SLOC \\ \hline \hline
		Location Share & Navigation & 384\\ \hline
		Calculator & Tools & 629 \\ \hline
		Pushup Buddy & Health \& Fitness & 965 \\ \hline
		SdbViewer & Tools & 1273 \\ \hline
		Cache Cleaner & Tools & 1493 \\ \hline
		JustCraigslist & Shopping & 3014 \\ \hline
		AltcoinPrices & Finance & 3069 \\ \hline
		Movie DB & Entertainment & 4727 \\ \hline
		BART Runner & Navigation & 6124 \\ \hline
		OINotepad & Productivity & 6220 \\ \hline
		Open Sudoku & Games & 6440 \\ \hline
		Mileage & Finance & 9931 \\ \hline
		Pedometer & Health \& Fitness & 13502\\ \hline
		FileDownloader & Tools & 17162 \\ \hline
		chanu & Social & 45856 \\ \hline
	\end{tabular}
\end{table}


 
 To generate tests that confirm to our criteria, we constructed test cases manually and used the sequences generated for our criteria to guide testing. For example, to exercise the \ee criterion, we move the phone to generate sensing events to trigger the callback sequences related to sensor event handlers followed by a GUI action.  Similarly,  for the \eaa criterion that involves two event callbacks and an API\_ASYNC,  we trigger these two events with different timing to generate different interleavings between the three callbacks. We run each app from 15 to 30 minutes, depending on the size of the app and stop when we believe all the callback sequences of the criteria are explored. As this paper focuses on defining and studying the effectiveness of the new coverage criteria, we leave the development of automatic generation of test cases to future work.
 
 As a comparison, we used one of the best Android testing tools Monkey~\cite{Choudhary:2015:ATI:2916135.2916273, Wang:2018:ESA:3238147.3240465}.  We adopted the settings done in the previous studies~\cite{sapienz, choudharytestingtools, wangguitesingtools}, used the default distribution of events and run each benchmark for 3 hours. 

We designed two experiments. In the first experiment, we ran Monkey and our testing and compared the number of bugs found, the time used to trigger the bugs and the coverage of our criteria achieved.  In the second experiment, for each of the approach, we record at every 5 minutes, the number of bugs (unique crash) triggered and the coverage of the \ee, \eas and \eaa criteria as well as the statement and GUI coverage achieved. Our goal is to observe whether the coverage of our criteria is increased before a bug is found. If we see the bugs often co-occur with the increases of our coverage, it suggests that testing guided by our criteria is useful for finding bugs. All the apps in our benchmark were run on a Motorola Nexus 6 with Android 7.1. In the following two sections, we report the results for each experiments respectively.

\subsection{Results of Our Testing Compared to Monkey}


In Table \ref{tab:faults}, we list the number of bugs found by our approach and Monkey. Our testing found a total of 17 bugs and Monkey found 14 bugs. Monkey and our testing reported 9 bugs in common (see Column {\it Both}). Although Monkey runs for hours and our testing runs in minutes, guided by our criteria, we are able to find 8 bugs that cannot be found by Monkey.  Monkey reported 5 bugs missed in our testing. Our inspection shows that 2 of these bugs are related to an \textit{Out of Memory} exception and 2 are related to the callbacks that react to network connectivity events. These bugs require to execute the apps for a long time, and we have not yet constructed such conditions in our testing. We foreseen that an automatic tool that fully explored our criteria would also trigger these cases. 

\begin{table}
	\centering
	\caption{Bugs found by the two approaches}\label{tab:faults}
	\begin{tabular}{|l|c|c|c|} \hline
		App & Both & Our testing & Monkey \\ \hline \hline
		Calculator & 4 & 0 & 0 \\ \hline
		AltcoinPrices & 1 & 0 & 0 \\ \hline
		Movie DB & 0 & 2 & 2 \\ \hline
		BART Runner & 0 & 1 & 2 \\ \hline
		OpenSudoku & 0 & 1 & 0 \\ \hline
		Mileage & 1 & 0 & 0 \\ \hline
		Pedometer & 2 & 2 & 0 \\ \hline
		FileDownloader & 0 & 1 & 0 \\ \hline
		chanu & 1 & 1 & 1 \\ \hline
		\textbf{Total} & 9 & 8 & 5 \\ \hline
	\end{tabular}
\end{table}


We compared the time used by Monkey and our testing to trigger each bug for the first time (a bug may trigger multiple times in testing). 
Since Monkey takes a long time to trigger the bugs that we did not find, we focus on comparing the bugs that were commonly discovered. In Table \ref{tab:time}, we report the average, minimum and maximum time used in seconds for the two approaches. We observe that by targeting \ee, \eas and \eaa, our testing is 7 times faster than Monkey to find the bugs on average. We also found that for bugs in apps such as Pedometer and chanu, Monkey took more than an hour to detect such bugs. The results demonstrated that our criteria not only help testing find bugs that other tools cannot find, but even for the bugs that other tools can find, we can find bugs faster.


\begin{table}
	\centering
	\caption{Compare the time used to trigger the common bugs}\label{tab:time}
	\begin{tabular}{|c|c|c|c|c|c|} \hline
		\multicolumn{3}{|c|}{Our testing} & \multicolumn{3}{|c|}{Monkey} \\ \cline{1-6} 
		Avg. (s) & Min (s) & Max (s) & Avg. (s)& Min (s) & Max (s) \\ \hline \hline
		318 & 4 & 1122 & 2187 & 2 & 10638 \\ \hline
	\end{tabular}
\end{table}

We also compare the coverage of our criteria achieved by the coverage guided testing and by Monkey. We want to investigate whether the black box testing tools like Monkey can achieve our criteria when we run the apps enough time. As shown in Table \ref{tab:coverage}, our approach achieved the better coverage on average. for all of the three criteria. For 12 out of 15 apps, our testing improved the coverage of \ee (Column EE), \eas (Column EAS) and \eaa (Column EAA) criteria or at least achieve the same coverage compared to Monkey.  The apps of {\it Location Share} and {\it Pushup Buddy} have reported the biggest improvement for the \ee criterion, as our testing is able to trigger the external events such as GPS location updates and sensor events while Monkey cannot. For the apps such as {\it BARTRunner} and {\it chanu}, Monkey was able to detect more bugs and achieve a better coverage, as these apps mostly consist of the GUI callbacks, which Monkey targets. Our testing still does not achieve 100\% coverage of our criteria, as some callback sequences are not feasible, and some conditions of triggering the sequences are hard to reason about by only inspecting the source code. 

\begin{table}
	\centering
	\caption{Coverage of our criteria achieved by the two approaches: the tabulars labeled with "-" indicate that there are no corresponding types of callbacks found in the app.}\label{tab:coverage}
	\begin{tabular}{|l|r|r|r|r|r|r|} \hline 
		\multirow{2}{*}{App} & \multicolumn{3}{|c|}{Our Testing} & \multicolumn{3}{|c|}{Monkey} \\ \cline{2-7} 
		& EE & EAS & EAA & EE & EAS & EAA \\ \hline \hline
		Location Share & 51.30 & - & - & 33.77 & - & - \\ \hline
		Calculator & 42.14 & - & - & 68.14 & - & - \\ \hline
		Pushup Buddy & 42.64 & 13.11 & 55.31 & 35.29 & 6.55 & 31.70 \\ \hline
		SdbViewer & 44.15 & - & 59.13 & 38.29 & - & 58.62 \\ \hline
		Cache Cleaner & 44.00 & 60.00 & 54.12 & 44.00 & 60.00 & 53.57 \\ \hline
		JustCraigslist & 36.46 & - & 26.72 & 34.92 & - & 24.42 \\ \hline
		AltcoinPrices & 29.40 & - & 59.40 & 30.10 & - & 63.31 \\ \hline
		Movie DB & 44.17 & 27.27 & 79.22 & 36.81 & 27.27 & 75.78 \\ \hline
		BARTRunner & 32.68 & - & 31.45 & 27.35 & - & 51.58 \\ \hline
		OI Notepad & 20.10 & 31.30 & - & 20.40 & 34.34 & - \\ \hline
		Open Sudoku & 45.78 & 25.45 & 76.58 & 24.64 & 7.27 & 60.00 \\ \hline
		Mileage & 25.09 & 23.25 & 43.24 & 23.58 & 34.88 & 21.81 \\ \hline
		Pedometer & 74.56 & 31.17 & 85.67 & 64.15 & 22.22 & 78.11 \\ \hline
		FileDownloader & 54.20 & 60.12 & 93.72 & 26.41 & 34.23 & 72.97 \\ \hline
		chanu & 30.21 & 27.18 & 40.57 & 35.41 & 28.64 & 48.18  \\ \hline
		\textbf{Average} & 44.01 & 36.09 & 63.62 & 38.80 & 30.97 & 57.78 \\ \hline
	\end{tabular}
\end{table}

\subsection{Correlations of the Bugs and the Increased Test Coverage}

\begin{figure*}
	\centering
	\begin{subfigure}[b]{0.49\textwidth}
		\centering
		\includegraphics[width=\textwidth]{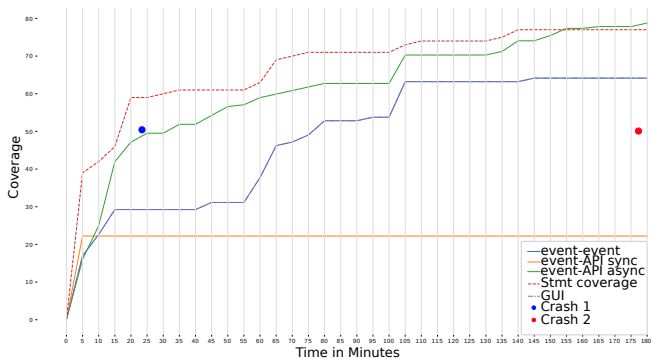}
		\caption{Coverage over time for Pedometer using Monkey\label{fig:pedometercoverage}}		
	\end{subfigure}
	\begin{subfigure}[b]{0.49\textwidth}
		\centering
		\includegraphics[width=\textwidth]{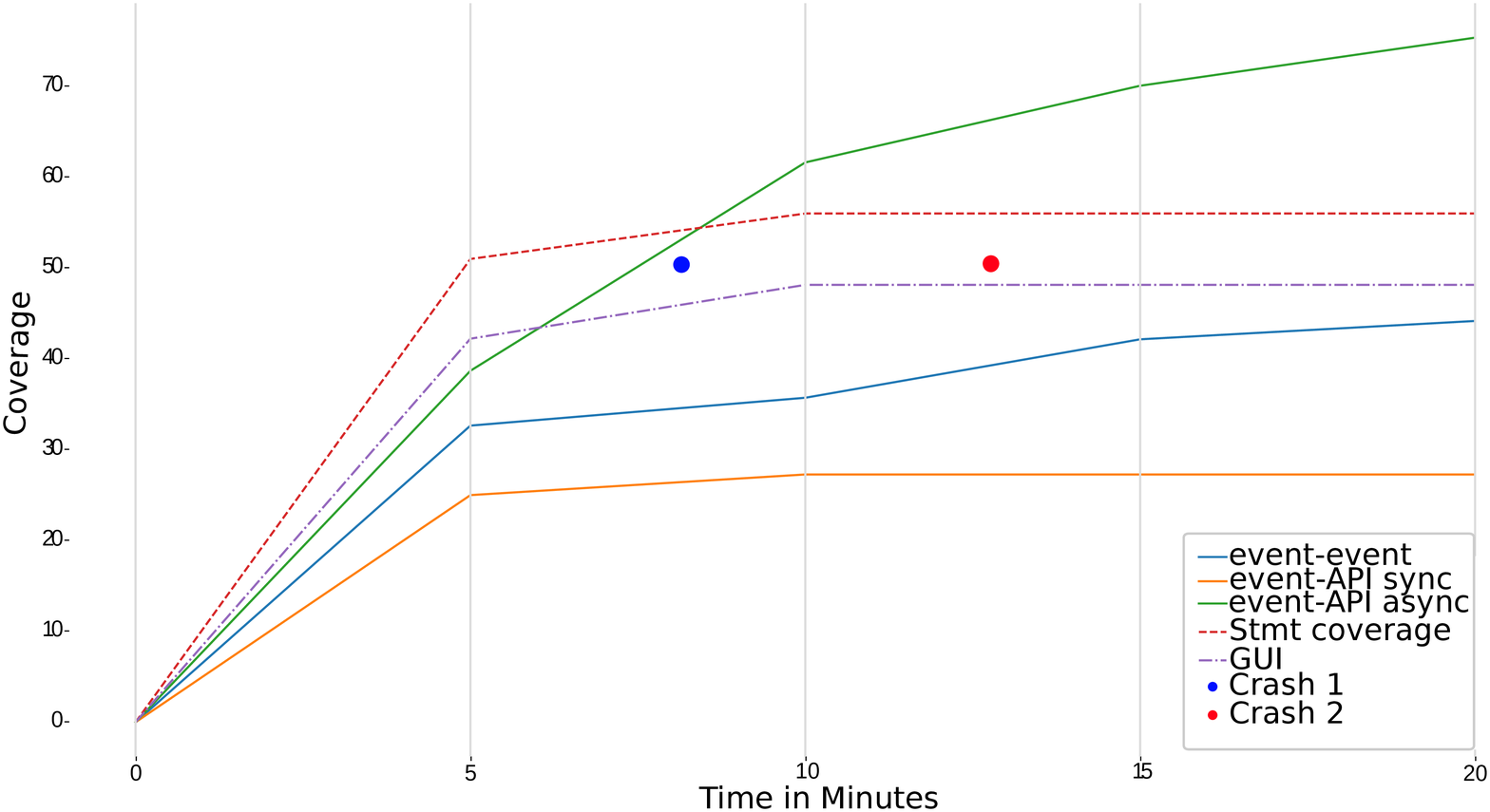}
		\caption{Coverage over time for Movie DB in our testing\label{fig:moviedbcoverage}}		
	\end{subfigure}
	\caption{Coverage over time for Pedometer and Movie DB\label{fig:coverage}}	
\end{figure*}

In Table \ref{tab:increase}, we report for all the 5-min periods, the number of bugs occurred with the increase of test coverage. As shown in Column {\it Bugs}, the \ee criterion is most effective for guiding testing, and we found 77\% of the bugs (24 bugs) occur with the increase of the coverage of this criterion, 16\% more (5 more bugs) than the GUI-based criterion, and 44\% more (14 more bugs) than the statement criterion. We found the \eas criterion is also very useful for finding bugs, as it correlates with the new bugs that the \ee and GUI criteria cannot find. We manually analyzed the root causes of all the bugs and confirmed that there are 8 race condition bugs. The occurrences of these bugs are all correlated with the increases of the \eaa criterion. Whereas, the increase of the GUI coverage is correlated with 3 of such 8 bugs and the statement coverage is correlated with 2 of these bugs. Therefore, the \eaa criterion is important for guiding testing to exercise the concurrency behaviors of Android apps. If we consider all of our three criteria together, we found that 30 out of 31 bugs are correlated with the increase of coverage of either the \ee, \eas or \eaa criterion.

\begin{table}
	\centering
	\caption {No. of bugs correlated with increased coverage }\label{tab:increase}
	\begin{tabular}{|l|c|} \hline
		Criterion  & Bugs \\ \hline \hline
		\ee & 24 \\ \hline
		\eas & 6  \\ \hline
		\eaa & 14 \\ \hline
		GUI & 19  \\ \hline
		Statement & 10  \\ \hline
		
	\end{tabular}
\end{table}

In Figures \ref{fig:pedometercoverage} and \ref{fig:moviedbcoverage}, we present two examples and show how the coverage for each criterion increased over every 5 minutes for the apps {\it Pedometer} and {\it Movie DB} respectively . We also marked the time when the bugs are first triggered (see the blue and red dots marked on the figures). We observe that within the 5-minute periods of the occurrences of the bugs in {\it Pedometer} (in Figure \ref{fig:pedometercoverage}, see the slots of 20-25 minutes that contains the blue dot, and 175-180 minutes that contains the red dot), the only coverage criterion increased is the \eaa criterion. Especially for the period of 175-180 minutes, the coverage of GUI and statements has not been updated for about 35 minutes. The developers who use the two criteria may already stop testing and miss the bug. Similarly for {\it Movie DB} (Figure \ref{fig:moviedbcoverage}), we observed that after 10-min testing, only the coverage of the \ee and \eaa criteria is updating. Thus, most likely, the sequences of these two criteria had triggered the second bug for {\it Movie DB}.

	\section{Related Work}\label{sec:relatedwork}

\paragraph{Coverage Criteria for Event-Driven Systems}
Memon \etal \cite{coveragecriteriagui} developed a family of coverage criteria for testing event-driven GUI applications. Their technique introduces a black-box model for event sequences to test permutations of events in GUI applications. For Android, most of the work on testing focus on generating GUI events \cite{guiripper, sapienz, swifthand, stoat, dynodroid, monkey} following similar techniques developed in \cite{coveragecriteriagui}.


Amalfitano \etal \cite{guiripper} present \textit{AndroidRipper} which instead of dynamically generates a GUI model (called ripping) and then generates test inputs, AndroidRipper systematically test the apps while doing the ripping. Mao \etal \cite{sapienz} developed SAPIENZ which combines random fuzzing and search-based exploration to maximize statement coverage. Su \etal \cite{stoat} introduces \textit{Stoat}, a tool that generates a weighted black-box GUI model dynamically and uses sampling to mutate the model to increase the result of an objective function. The objective function considers the model coverage, statement coverage and model diversity (how did the GUI model change). 

All these tools focus on some way to increase statement coverage or the coverage of a black-box GUI model. In this paper, we focus on designing new coverage criteria for testing Android apps and show that the GUI models and statement coverage lack important information that needs to be tested. 

\paragraph{Static Models for Testing}
In this paper, we use a static model, CCFA, to generate callback sequences for our criteria. Similarly, Azim and Neamtiu \cite{a3e} implemented a technique (in the tool $\mathit{A^3E}$) that generates a control flow graph that contains legal transitions between Activities. They also developed a targeted exploration technique to guide tests to improve coverage on their control flow graph. Yang \etal \cite{Yang:2015:SWT} developed \textit{Windows Transition Graphs} (WTGs) to model sequences windows and GUI events. They then generate GUI events by traversing their WTGs. The WTGs have been also used for testing resource leaks \cite{wtgleaks}. Both of these approaches focus just on GUI behaviors. The CCFA covers GUI behaviors and also includes other external events (such as camera or sensors) and callbacks invoked in API methods. Neither $\mathit{A^3E}$ and WTGs cover these behaviors. 

\paragraph{Testing for Concurrency}
For testing concurrent systems, similar techniques to our work have been used for different systems. Deng \etal \cite{dengconcurrency} and Choudhary \etal \cite{choudharyconcurrency} used pair of concurrent functions as a coverage metric for testing C/C++ applications and thread-safe Java classes respectively. The former work uses the coverage metric to select the pre-defined inputs whereas the latter uses the metric for input generation. Similarly, Tasharofi \etal \cite{tasharoficoncurrency} developed different coverage criterion based on pairs of concurrent operations for actor programs. This related work shows that pairs of functions are an effective metric as coverage criteria for testing concurrent systems. 

To help detect concurrency issues in Android, most of the related work used happen-before relation on dynamic traces. Hsiao \etal \cite{raceandroidhsiao} and Maiya \etal \cite{raceandroidmaiya} both developed concurrent models and happen-before relation for Android to detect race conditions between callbacks. \cite{racebielik} developed new techniques for scaling the inference of happens-before relations. All these techniques depend on dynamic traces generated from testing. Contrary, the callback sequences of our \eaa criterion is generated from the CCFA which is a model based on the source code of the app. Li \etal \cite{litesting} present a similar technique to ours to detect concurrency bugs between GUI callback listeners. They generate input events to test the interactions of these callbacks. Our \eaa criterion focuses on interleavings related to asynchronous callbacks invoked in API methods. 

	\section{Conclusions and Future Work}
This paper introduced three white-box coverage criteria based on callback sequences. The \ee criterion aims to cover callback sequences from external events, including GUI, sensing and other types of events. The \eas criterion requires to test different behaviors from synchronous callbacks invoked in API methods. The \eaa is to check concurrent behaviors between external events and asynchronous callbacks in API methods. Our evaluation results show that testing guided by our criteria can find new bugs other testing tools cannot and found bugs 7 times faster than Monkey. We also demonstrated that there are correlations between the occurrence of the bugs and the increases of the coverage of our criteria. For future work, we plan to design new input generation tools or augment existing tools based on our criteria. For example, a tool such as \textit{Stoat} \cite{stoat} can use our criteria to improve their objective function when generating new inputs.



	\bibliographystyle{IEEEtran}
	\bibliography{bibliography}
	
\end{document}